\documentstyle[12pt]{article}

\parskip=\medskipamount            % space between paragraphs (LaTeX)
\def\ket#1{| #1\rangle}                 % | >
\def\7#1#2{\mathop{\null#2}\limits^{#1}}        % puts #1 atop #2

\def\beee{\begin{equation}}
\def\eeee{\end{equation}}
\def\dggg{^{\dagger}}
\begin{document}

\bibliographystyle{unsrt}

\begin{center}
{\Large \bf CONSTRUCTION OF BOSONS AND FERMIONS OUT OF QUONS}
\\[.1in]

O.W. Greenberg\footnote{email address, owgreen@physics.umd.edu}\\
{Center for Theoretical Physics\\
Department of Physics \\
University of Maryland\\
College Park, MD~~20742-4111}\\

and\\

J.D. Delgado\footnote{email address, jdelgado@wam.umd.edu}\\
{Center for Theoretical Physics\\
Department of Physics\\
University of Maryland\\
College Park, MD~~20742-4111}

UMPP 01-064\\

\end{center}

\vspace{5mm}

\begin{abstract}
The quon algebra describes particles, ``quons,'' that are neither
fermions nor bosons, using a label $q$ that parametrizes a smooth
interpolation between bosons ($q = 1$) and fermions ($q = -1$).
Understanding the relation of quons on the one side and bosons or
fermions on the other can shed light on the different properties
of these two kinds of operators and the statistics which they
carry.  In particular, local bilinear observables can be
constructed from bosons and fermions, but not from quons.  In
this paper we construct bosons and fermions from quon operators.
For bosons, our construction works for $-1 \leq q \leq 1$.  The
case $q=-1$ is paradoxical, since that case makes a boson out of
fermions, which would seem to be impossible.  None the less, when
the limit $q \rightarrow -1$ is taken from above, the
construction works. For fermions, the analogous construction
works for $-1 \leq q \leq 1$, which includes the paradoxical case
$q=1$.
\end{abstract}

%\begin{center}

\section{Introduction}
Why are all identical particles in nature either bosons or
fermions? Attempts to answer this question have gone in two
directions: (1) to study other possible types of quantum
statistics theoretically and to propose and analyze experimental
searches for violations of bose or fermi statistics \cite{owgrnm},
(2) to find theoretical arguments that only bose or fermi
statistics occur in nature and thereby understand better why only
bose and fermi statistics can occur \cite{ckcowg}.  The book
edited by R.C. Hilborn and G.M. Tino \cite{hiltin} has the
proceedings of a recent conference on statistics.

A better understanding of the relation of bose and fermi
statistics to other possible statistics should be a useful tool
for analysis of both these issues.  In this letter we study the
construction of bose and of fermi operators using quons
\cite{owg91}, a type of particle statistics that interpolates
smoothly between bose and fermi statistics and can provide a
description of particles that violate bose or fermi statistics by
a small amount\cite{statexp}.  We give ansatzes for the
construction of bosons from quons with $q$ in the range $-1 \leq
q \leq 1$ and for the construction of fermions from quons in the
range $-1 \leq q \leq 1$. In each case the ansatzes represent
bose or fermi creation or annihilation operators in terms of an
infinite degree series of quon operators.  Surprisingly, our
constructions are valid for bosons represented by an infinite
series of products of fermions as well as for fermions
represented by an infinite series of bosons.

The bose and fermi algebras are
\beee
[b_k,b\dggg_l]_-=\delta_{k,l}
\eeee
\beee
[f_k,f\dggg_l]_+=\delta_{k,l},
\eeee
and the Fock representation that we will always consider is given
by
\beee
b_k \ket{0}=0,~~ {\rm or} ~~f_k \ket{0}=0.
\eeee
The quon algebra is
\beee
a_k a_l \dggg-q a_l \dggg a_k=\delta_{kl}.               \label{q}
\eeee
We consider the Fock-like representation with the vacuum condition
\beee
a_k |0\rangle=0.                                         \label{f}
\eeee

These two conditions determine all vacuum matrix element of polynomials in the
creation and annihilation operators.  In the case of free quons, all
non-vanishing vacuum matrix elements have the same number of annihilators
and creators.  For such a matrix element with all annihilators to the left and
creators to the right, the matrix element is a sum of products of
``contractions'' of the form $\langle 0|a a\dggg |0 \rangle$ just as in the case
of bosons and fermions.  The only difference is that the terms are multiplied by
integer powers of $q$.  The power can be given as a graphical rule:  Put
$\circ$'s
for each annihilator and $\times$'s for each creator in the order in which
they occur in the matrix element on the x-axis.
Draw lines above the x-axis connecting the pairs that are
contracted.  The minimum number of times these lines cross
is the power of $q$ for that term in the matrix element \cite{owg91}.
For $q=\pm 1$ this agrees with the usual rule for bosons and fermions.

The physical significance of $q$ for small violations of fermi
statistics is that $q=2 v_f -1$, where the parameter $v_f$ gives
the deviation of the two-particle density matrix from fermi
statistics, \beee \rho_2=(1-v_f) \rho_a+v_f \rho_s, \eeee where
$\rho_{a(s)}$ is the antisymmetric (symmetric) two-particle
density matrix. For small violations of bose statistics, the
two-particle density matrix is \beee \rho_2=(1-v_b) \rho_s+v_b
\rho_a. \eeee For this case $q=1- 2v_b$.

For $q$ in the open interval $(-1, 1)$ all representations of the
symmetric group occur.  As $q \rightarrow 1$, the symmetric
representations are more heavily weighted and at $q=1$ only the
totally symmetric representation remains; correspondingly, as $q
\rightarrow -1$, the antisymmetric representations are more
heavily weighted and at $q=-1$ only the totally antisymmetric
representation remains.  Thus for a general $n$-quon state, there
are $n!$ linearly independent states for $-1<q<1$, but there is
only one state for $q= \pm 1$. We emphasize something that many
people find very strange: {\it there is no commutation relation
between two creation or between two annihilation operators,}
except for $q= \pm 1$, which, of course, correspond to bose and
fermi statistics.  Indeed, the fact that the general $n$-particle
state with different quantum numbers for all the particles has
$n!$ linearly independent states proves that there is no such
commutation relation between any number of creation (or
annihilation) operators. An even stronger statement holds:  There
is no two-sided ideal containing a term with only creation
operators.

Quons are an operator realization of ``infinite statistics'' which were found as
a possible statistics by Doplicher, Haag and Roberts \cite{dhr} in their general
classification of particle statistics.  The simplest case, $q=0$, \cite{owg90}
suggested to one of us (OWG) by Hegstrom,
was discussed earlier in the context of operator algebras by Cuntz \cite{cun}.

It is amusing that, even though quon fields are not local,
the free quon field obeys the TCP
theorem and Wick's theorem holds for quon fields \cite{owg91}.

We consider three ways to construct bose or fermi creation and annihilation
operators using quons.  The first way, in Sec. 2, uses a general ansatz for
the bose or fermi operators in terms of an infinite degree series of normal
ordered products of quon operators with undetermined coefficients.  Direct
substitution of the ansatz in the bose or fermi commutation relation yields
conditions that determine the coefficients.  We show that the number of
constraints equals the number of coefficients, but we do not give a formula
for the general term using this method.  The second way, in Sec. 3, uses
the same ansatz, but determines the coefficients by requiring the validity
of the bose or fermi permutation symmetry relation for states with a given
number of
quanta. As the number of quanta is increased, the higher coefficients are
determined.  Again we do not give a formula for the general term.
The third approach, in Sec. 4,
is to expand the bose operator in terms
that act on only one sector with a given number of quanta; that is to expand
using terms that have projection operators on the subspace with a given
number of quanta of the quons, or, in other words, to isolate
the terms that contribute for a given number of quanta.  This third
approach yields simple explicit formulas for an arbitrary term in the
infinite series representation of the bose operators in terms of quon
operators.  The corresponding results for the representation of fermi
operators in terms of quon operators are analogous to the results for
bose operators.

We were surprised to find that our results hold for the expansion
of a bose operator using fermi operators ($q=1$) and for the
expansion of a fermi operator using bose operators ($q=-1$).  We
discuss the relevant limiting cases in Sec. 5.

\section{Direct calculation in the algebra}

Choose the $a_k$ and $a_l\dggg$ operators to obey the quon relations
(\ref{q})
in the Fock-like representation (\ref{f}). Expand a bose operator $b_k$ as
\beee
b_k=c_{kl} a_l + c_{klmn}a_l\dggg a_m a_n +
%a_{s1}\dggg a_{s2}\dggg (a_{s2}a_{s1}a_k+a_{s2}a_ka_{s1}+a_ka_{s2}a_{s1}
%+a_{s1}a_{s2}a_k+a_{s1}a_ka_{s2}+a_ka_{s1}a_{s2})
%+ \cdots +\sum_{j=1}^{n-1}\sum_P c_{sP1,sP2, \cdots sPn}a_{sP1}a_{sP2}
\cdots.
\eeee
%with $a_{sn}\equiv a_k$, where the permutation $P$ takes $1 \rightarrow P1$,
%$2 \rightarrow P2$, etc and summation over repeated indices $si$ is understood.
Unless stated otherwise, repeated indices are summed over.
The expression for $a_k\dggg$ is always the adjoint of the expression for
$a_k$.  We want to find the
simplest ansatz that will allow the construction.  Translation and
rotation invariance are compatible with the following much simpler ansatz,
\beee
b_k = a_k + x  a_t\dggg a_k a_t +y  a_t\dggg a_t a_k + \cdots.   \label{2}
\eeee
The terms in (\ref{2}) suffice to give all terms of the form $a\dggg a$ in the
commutator $[b\dggg_l,b_k]_-$.  There are two different types of terms bilinear
in $a\dggg$ and $a$, $a\dggg_l a_k$ and $\delta_{kl} a\dggg_s a_s$; these lead to
two equations that determine $x$ and $y$,
\beee
q(x^2+y^2) +2xy+2(x+qy) +q-1=0,
\eeee
\beee
x^2+y^2+2qxy+2(qx+y)=0.
\eeee
The solutions are
\beee
x=\pm \frac{1}{\sqrt{2(1+q)}},~~y=\pm \frac{1}{\sqrt{2(1+q)}} -1. \label{sol}
\eeee
Inserting these solutions in (\ref{2}), we find
\beee
b\dggg_l=a\dggg_l + \pm \frac{1}{\sqrt{2(1+q)}}a_t\dggg a\dggg_l a_t+
(\pm \frac{1}{\sqrt{2(1+q)}}-1)a\dggg_la\dggg_ta_t + \cdots,
\eeee
where the result is valid up to terms in the commutator with four operators.
For the upper choice of sign the result reduces to $b\dggg_l=a\dggg_l$ when
$q=1$, as we expect.  We disfavor the lower choice of sign, which does not
so reduce, because it gives an unwanted negative sign for the two-particle
state even though it gives a symmetric two particle state for $q=1$,
\beee
b\dggg_{l2} b\dggg_{l1}\ket{0}=-a\dggg_{l2} a\dggg_{l1}\ket{0}.
\eeee

Similarly, for the fermi case, we can expand the fermi operator
as: \beee f_k = c_{kl} a_l + c_{klmn}a_l\dggg a_m a_n + \cdots.
\eeee which can be written more simply as: \beee f_k = a_k + x'
a_t\dggg a_k a_t + y'  a_t\dggg a_t a_k + \cdots. \eeee Now we
consider the anticommutator $[f_l\dggg,f_k]_+$, which gives us
the following set of equations: \beee q(x'^2 + y'^2) + 2x'y' +
2(x' + qy') + q + 1=0, \eeee \beee x'^2 + y'^2 + 2qx'y' + 2(qx' +
y')=0. \eeee which yields \beee x' = \pm \frac{1} {\sqrt{2(1 -
q)}}, ~~y' = \mp \frac{1} {\sqrt{2(1 - q)}} - 1. \label{sol1}
\eeee Inserting the solutions the fermi operator becomes 
\beee
f\dggg_l=a\dggg_l + \pm \frac{1}{\sqrt{2(1-q)}}a_t\dggg a\dggg_l
a_t+ (\mp \frac{1}{\sqrt{2(1-q)}}-1)a\dggg_l a\dggg_t a_t +
\cdots. 
\eeee 
Here the lower choice of sign reduces the result to
$f\dggg_l=a\dggg_l$ and the upper choice gives the two-particle
state with an unwanted extra minus sign.

The general ansatz for $b_k$ will be an infinite series with
terms of the form $a^{\dagger~n-1} a^n$.  Taking account of the
possibility of renaming indices, there can be $n!$ different
terms of this type, since the creation operators can be labeled
by indices $t1, t2, \cdots, t(n-1)$ in fixed order and the
annihilation operators can be labeled by $k, t1, t2, \cdots,
t(n-1)$ in $n!$ different orders, each such term having an
independent coefficient.  The commutator $[b_k,b\dggg_l]_-$ will
have terms of the form $a^{\dagger~n-1} a^{n-1}$ of two types,
those with a factor $\delta_{k,l}$ and those in which the indices
$k$ and $l$ appear on operators in the product of annihilation
and creation operators.  There are $(n-1)!$ independent terms of
the first kind, since the indices $k1, k2, \cdots, k(n-1)$ can be
fixed for--say--the creation operators and the indices can be
permuted in $(n-1)!$ ways in the annihilation operators. To count
the number of different terms of the second type, fix the order
of $a\dggg_{t1}  a\dggg_{t2} \cdots a\dggg_{t(n-2)}$. The
creation operator $a\dggg_l$ can be put in $(n-1)$ places among
the $a\dggg$'s. The annihilation operators $a_{t1} a_{t2} \cdots
a_{t(n-2)}$ can be permuted in $(n-2)!$ ways. The annihilation
operator $a_k$ can be put in $(n-1)$ places. Thus the number of
different terms of the second type is
$(n-1)(n-2)!(n-1)=(n-1)(n-1)!$ and the total number of different
terms is $(n-1)!+(n-1)(n-1)!=n!$ The equality of the number of
independent coefficients and the number of independent
constraints makes it plausible that a solution exists for an
arbitrary term in the infinite series representation of the bose
operators in normal-ordered products of the quon operators.

\section{Direct calculation on the states}

A representation of the bose operators in terms of quon operators will have
the property that a product of $n$ bose creation operators acting on the
vacuum will be symmetric under permutations of the bose creation operators.
When the bose creation operators are represented by quon operators, the
$n$ particle state will be symmetric under permutations of the quon creation
operators.  We can use this property to simplify the expansion of the
bose operators in terms of the quon operators.

For the case of one quantum, we clearly need
\beee
b\dggg_k=a\dggg_k.
\eeee
For the case of two quanta, we can take
\beee
b_k = a_k + x  a_t\dggg a_k a_t +y  a_t\dggg a_t a_k
\eeee
as before.  Direct calculation gives
\beee
b\dggg_{k2} b\dggg_{k1} \ket{0} =
[(1+y) a\dggg_{k2} a\dggg_{k1} + x a\dggg_{k1}a\dggg_{k2}]\ket{0}.
\eeee
We require
\beee
b\dggg_{k2} b\dggg_{k1} \ket{0} =
{\cal N}(2;q) [a\dggg_{k2} a\dggg_{k1}+a\dggg_{k1}a\dggg_{k2}]\ket{0}
\eeee
in order to enforce bose symmetry on the two-particle state.
Using \cite{owg91}
\beee
{\cal N}(n;q) = \frac{1}{\sqrt{n! [n]_q!}},~~[n]_q=1+q+ \cdots +q^{n-1},~~
[n]_q!=[1]_q[2]_q \cdots [n]_q,
\eeee
we find ${\cal N}(2;q)=1/\sqrt{2(1+q)}$ which yields the same
solution for $x$ and $y$ found in (\ref{sol}).

We can rewrite these solutions as \beee b\dggg_l=a\dggg_l
(1-a\dggg_t a_t) \pm \frac{1}{\sqrt{2(1+q)}} (a\dggg_t a\dggg_l +
a\dggg_l a\dggg_t)a_t.    \label{b1} \eeee When we act with
$b\dggg_{l2}$ as just defined on the single-particle state
$b\dggg_{l1} \ket{0} = a\dggg_{l1} \ket{0}$ we find that the
first term in (\ref{b1}) annihilates the single-particle state
and the second term, acting on the single-particle state, creates
the correctly normalized symmetric two-particle state.  This
suggests a simple strategy to construct the general form of the
bose creation operator: for the term in the creation operator that
is to act on the $(n-1)$-particle state and create the properly
normalized symmetric $n$-particle state, divide the term into two
parts, $b^{(n)\dagger}_l \equiv b^{1(n)\dagger}_l +
b^{2(n)\dagger}_l$, make the first part annihilate the
$(n-1)$-particle state and the second part create the normalized
symmetric $n$-particle state.  We only require the bose
commutation relation to be satisfied on the Fock space of bose
operators.  Thus we can assume that the states created by $n-1$
bose creation operators expressed in terms of the quon creation
operators acting on the vacuum are totally symmetric and
normalized; i.e., have the form \beee b\dggg_{l1} b\dggg_{l2}
\cdots b\dggg_{l(n-1)} \ket{0} = {\cal N}(n-1;q) \sum_P
a\dggg_{lP1} a\dggg_{lP2} \cdots a\dggg_{lP(n-1)} \ket{0}. \eeee
The second part, $b^{2(n)\dagger}_l$ can immediately be written
down. The $n-1$ annihilation operators standing to the right can
be written in a fixed order, since they will be symmetrized
automatically when they contract with the $n-1$ symmetrized
creation operators that act on the vacuum to create the
$n-1$-particle symmetric state. The summed over creation
operators standing to the left can also be written in a fixed
order; they will be symmetrized because they are summed over with
the symmetrized annihilation operators.  The only creation
operator that must be moved around is the creation operator
$a\dggg_l$ that carries the index of the bose operator $b\dggg_l$
that we are constructing.  The result is
\begin{eqnarray}
b^{2(n)\dagger}_l&=&{\cal M}(n;q)[a\dggg_l a\dggg_{t1} a\dggg_{t2}
\cdots a\dggg_{t(n-1)} + a\dggg_{t1} a\dggg_l a\dggg_{t2} \cdots a\dggg_{t(n-1)}
\nonumber \\
& & + \cdots + a\dggg_{t1} a\dggg_{t2} \cdots a\dggg_{t(n-1)} a\dggg_l]
 a_{t(n-1)}\cdots a_{t2}a_{t1},                              \label{bn}
\end{eqnarray}
\beee
{\cal M}(n;q)=\frac{1}{n [n]_q! [n-1]_q!}.
\eeee
(To construct $b^{2(n)\dagger}_l$ that can be used to represent the bose operators
on an arbitrary quon state, replace the product of $a$'s in (\ref{bn}) by
$\sum_P a_{tP1} a_{tP2} \cdots a_{tP(n-1)}/[n-1]_q!$)
The form of the first part, $b^{1(n)\dagger}_l$ is more complicated,
because the term that
creates the $n$-particle state acting on the state with $n-1$ particles
will do complicated things acting on states with more than $n$ particles.
We can avoid this problem and gain a simple result if we use projection
operators so that the term in $b\dggg$ that makes an $n$-particle state
contributes only when it acts on the $n-1$-particle state.

\section{Calculation using projection operators}

Define the projection operator $\Lambda_n$ to give one acting on an
$n$-particle state and zero otherwise.  We construct $\Lambda_n$ by
\beee
\Lambda_n= \frac{sin~ \pi (N-n)}{\pi (N-n)},
\eeee
where the explicit formula for $N$ in terms of the quon operators
has been given by S. Stanciu \cite{sta}.
Using the results of the previous section the formula for $b\dggg_l$ is
\beee
b\dggg_l=\sum_{n=1}^{\infty}\Lambda_n b_l^{2(n) \dagger} \Lambda_{n-1}
\label{result}
\eeee
Equation (\ref{result}) is the desired formula for the bose creation
operator expressed in terms of the quon operators.

The analogous formula for constructing fermions from quons is
\beee
f\dggg_l=\sum_{n=1}^{\infty}\Lambda_n f_l^{2(n) \dagger} \Lambda_{n-1},
\label{result2}
\eeee
\begin{eqnarray}
f^{2(n)\dagger}_l & = & {\cal M}(n;-q)[a\dggg_l a\dggg_{t1} a\dggg_{t2}
\cdots a\dggg_{t(n-1)} - a\dggg_{t1} a\dggg_l a\dggg_{t2} \cdots
a\dggg_{t(n-1)}                                      \nonumber \\
& & + \cdots \pm a\dggg_{t1} a\dggg_{t2} \cdots a\dggg_{t(n-1)} a\dggg_l]
 a_{t(n-1)}\cdots a_{t2}a_{t1},
\end{eqnarray}
\beee
{\cal M}(n;-q)=\frac{1}{n [n]_{-q}! [n-1]_{-q}!}.
\eeee

\section{Bosons in terms of fermions and fermions in terms of bosons}

Consider the solution (\ref{b1}) for the first nontrivial term in the expansion
of the bose operator in terms of quons.
The limiting case $q=-1$ seems to be ill-defined
because of the factor $\sqrt{2(1+q)}$ in the denominator; however the numerator
has the factor $(a\dggg_t a\dggg_l + a\dggg_l a\dggg_t)$ which also vanishes
for $q \rightarrow -1$.  To evaluate the limit for $q \rightarrow -1$ we
calculate the matrix element
\beee
(b\dggg_{l2} b\dggg_{l1}\ket{0}, b\dggg_{k2} b\dggg_{k1}\ket{0}).
\eeee
The result
\beee
\delta_{l1,k1} \delta_{l2,k2} + \delta_{l1,k2} \delta_{l2,k1}
\eeee
is valid for all $q$ and shows that for two-particle states
the bose operator is correctly represented
by the quon series in the limit $q \rightarrow -1$.  Analogous results hold for
the general case for both the bose and fermi operators.

It has been known for a long time \cite{jor} that fermions can be
represented by bose
operators in two spacetime dimensions.  The present construction holds in any
number of dimensions.

\flushleft{{\large\bf Acknowledgement}}

J.D. Delgado would like to thank S. Garzon for helpful discussions.

\end{document}